\documentclass[aps,prd,twocolumn,showpacs,groupedaddress]{revtex4}

\usepackage{graphicx}
\usepackage{color}
\usepackage{slashed}
\usepackage{float}

\def\apj #1 #2 #3 {#1, ApJ, {\bf #2}, #3}
\def\apjl #1 #2 #3 {#1, ApJ, {\bf #2}, L#3}
\def\apjs #1 #2 #3 {#1, ApJS, {\bf #2}, #3}
\def\aap  #1 #2 #3 {#1, A\&A, {\bf #2}, #3}
\def\mnras #1 #2 #3 {#1, MNRAS, {\bf #2}, #3}
\def\pra #1 #2 #3 {#1, Phys.~Rev.~A., {\bf #2}, #3}
\def\prb #1 #2 #3 {#1, Phys.~Rev.~B., {\bf #2}, #3}
\def\prc #1 #2 #3 {#1, Phys.~Rev.~C., {\bf #2}, #3}
\def\prd #1 #2 #3 {#1, Phys.~Rev.~D., {\bf #2}, #3}
\def\pre #1 #2 #3 {#1, Phys.~Rev.~E., {\bf #2}, #3}
\def\prl #1 #2 #3 {#1, Phys.~Rev.~Lett., {\bf #2}, #3}
\def\plb #1 #2 #3 {#1, Phys.~Lett.~B., {\bf #2}, #3}
\def\science #1 #2 #3 {#1, Science., {\bf #2}, #3}
\def\nature #1 #2 #3 {#1, Nature., {\bf #2}, #3}
\def\nphysa #1 #2 #3 {#1, Nucl.~Phys.~A., {\bf #2}, #3}
\def\nphysb #1 #2 #3 {#1, Nucl.~Phys.~B., {\bf #2}, #3}
\def\nphysbs #1 #2 #3 {#1, Nucl.~Phys.~B.~Suppl., {\bf #2}, #3}

\def\h#1{\hbox{${}^{#1}$H}}
\def\he#1{\hbox{${}^{#1}$He}}

\def\be#1{\hbox{${}^{#1}$Be}}

\def\h502{\hbox{$ h^{2}_{50}$}}

\def\fun#1#2{\lower3.6pt\vbox{\baselineskip0pt\lineskip.9pt
  \ialign{$\mathsurround=0pt#1\hfil##\hfil$\crcr#2\crcr\sim\crcr}}}
\begin{document}

\title{New observational limits on dark radiation in brane-world cosmology}

\author{Nishanth Sasankan}
\email[]{nsasanka@nd.edu}
\affiliation{Center for Astrophysics, Department of Physics, University of Notre Dame, Notre Dame, IN 46556}

\author{Mayukh Raj Gangopadhyay}
\email[]{mgangopa@nd.edu}
\affiliation{Center for Astrophysics, Department of Physics, University of Notre Dame, Notre Dame, IN 46556}

\author{Grant J Mathews}
\email[]{gmathews@nd.edu}
\affiliation{Center for Astrophysics, Department of Physics, University of Notre Dame, Notre Dame, IN 46556}

\author{Motohiko Kusakabe}
\email[]{mkusakab@nd.edu}
\affiliation{Center for Astrophysics, Department of Physics, University of Notre Dame, Notre Dame, IN 46556}


\date{\today} 

\begin{abstract}
A dark radiation term arises as a correction to the energy momentum tensor in the simplest five-dimensional RS-II brane-world cosmology. In this paper we revisit the constraints on dark radiation based upon the newest results for   light-element nuclear reaction rates, observed light-element abundances and the power spectrum of the Cosmic Microwave Background (CMB). Adding dark radiation during big bang nucleosynthesis alters the Friedmann expansion rate causing the nuclear reactions to freeze out at a different temperature. This  changes the final light element abundances at the end of BBN. Its influence on the CMB is to change the effective expansion rate at the surface of last scattering.  We find that  our adopted  BBN constraints reduce   the allowed range for dark radiation to  between  $-12.1\%$ and $+6.2\%$ of the ambient  background energy density. Combining this result with fits to the CMB power spectrum, the range decreases to $-6.0\%$ to $+6.2\%$. Thus,  we find, that the ratio of dark radiation to the background total relativistic mass energy density $\rho_{\rm DR}/\rho$ is consistent with zero although in the BBN analysis there could be a slight preference for a negative contribution. However, the BBN constraint depends strongly upon the adopted primordial helium abundance. 
\end{abstract}

\maketitle


\section{INTRODUCTION}

One of the proposed solutions to the hierarchy problem among the fundamental forces is the introduction of compact extra dimensions. However, this creates a new hierarchy problem between the weak forces and the size of the compact extra dimensions. A possible solution was suggested by Randall and Sundrum \cite{Randall} by introducing a non-compact large extra dimension. In that model, the observed  universe is a four-dimensional spacetime embedded in a five-dimensional anti-de-sitter space (AdS5).

The projected  three-space Friedmann equation of the five dimensional universe reduces to \cite{Langlois}:
\begin{equation}
\left(\frac{\dot{a}}{a}\right)^2
=\frac{8 \pi G_{\rm N}}{3} \rho
-\frac{K}{a^2}+\frac{\Lambda_{4}}{3}
+\frac{\kappa_{5}^4}{36}\rho^2 + \frac{\mu}{a^4}~~.
\label{Friedmann}
\end{equation}
Here $a(t)$ is the usual scale factor for the three-space at time $t$, while $\rho$ is the energy density of matter in the normal three space. $G_N$ is the four dimensional gravitational constant and is related to its  five dimensional counterpart $\kappa_5$ by
\begin{equation}
G_{\rm N} = \kappa_{5}^4 \lambda / 48 \pi~~,
\end{equation}
where $\lambda$ is the intrinsic tension on the brane and $\kappa_5^{2}= M_5^{-3}$, with $M_5$ the five dimensional Planck mass. The $\Lambda_4$ in the third term on the right-hand side is the four dimensional cosmological constant and is related to its  five dimensional counterpart by 
\begin{equation}
\Lambda_{4}=\kappa_{5}^4 \lambda^2 /12 + 3 \Lambda_{5}/4~~.
\end{equation}
Note that for $\Lambda_{4}$ to be close to zero, $\Lambda_{5}$ should be negative. Hence the spacetime is AdS5. 

In standard Friedmann cosmology only the first three terms arise. The fourth term is probably negligible during most of the radiation dominated epoch since $\rho^{2}$ decays as $a^{-8}$ in the early universe. However this term could be significant during the beginning of the  epoch of inflation \cite{Maartins00, Okada16,Mayukh}. 

The last term is the dark radiation \cite{Binetruy00, Mukohyama00}. It is called radiation since it scales as ${a^{-4}}$. 
It is a constant of integration that arises from  the projected Weyl tensor describing the effect of graviton degrees of freedom on the dynamics of the brane.  
One can think of it, therefore, as  a projection of the curvature in higher dimensions.  In principle it could be either positive or negative. 

Although it is dubbed  dark radiation it is not related to  relativistic particles. Since it does not gravitate, flow or scatter as would a light neutrino   species, its effects on the cosmic microwave background (CMB) is different than that of normal radiation. Nevertheless, since it scales like radiation, its presence can alter the expansion rate during the radiation dominated epoch. This effect has been studied previously by several authors \cite{Ichiki, Bratt}. Here we update those previous studies in the context of newer constraints on light elemental abundances, BBN nuclear reaction rates and the CMB.

\section{Effects of Dark Radiation}
Dark radiation can have significant effects on the light element abundances produced during big bang nucleosynthesis (BBN).  It also affects the angular power spectrum of the CMB. Altering the expansion rate changes the temperature at which various nuclear reactions freeze out. This leads to  deviations in the final BBN light element abundances. In this paper we define $\rho_{DR} \equiv  (3 /8 \pi G_{\rm N})\mu/a^4$  as the energy density of the dark radiation, and parametrize $\rho_{DR} /\rho$ to be the ratio of the energy density of the dark radiation to the total energy density in relativistic particles at 10 MeV (before $e^+-e^-$ annihilation). The corresponding changes in the BBN abundances and the CMB power spectrum are then computed. 

Observations of the CMB and the Hubble expansion rate $H_0$ suggests the possible  existence of an additional density in the form of dark radiation \cite{cheng, calabrese}. The effect of the altered expansion rate on BBN was first discussed by \cite{Hoyle}. This effect was further studied by many authors \cite{shvartsman, Peebles, steigman1977}. We note, however, that exotic relativistic particles that do not interact with normal background particles have also been referred to as dark radiation.  However,  they are not the same as the dark radiation discussed here. The effects of these exotic particles have been studied by numerous authors \cite{justin, steigman2012, Nollett2014}. 

The effect of an altered expansion rate during the epoch of BBN and CMB has been studied in the context of constraining the effective number of neutrino species $N_{eff}$\cite{bbnreview, cicoli, steigman1977, boesgaard, steigman, Nollett2011, Hamann, Barger,Kholpov,Doroshkevich}. Positive and negative dark radiation can be associated with the uncertainty in the number of neutrino species $\Delta N_{\nu}$. The standard model suggests that we have 3 types of neutrinos, therefore we assume this to be true during the epoch of BBN. An addition of dark radiation ($\rho_{DR}$) can be related to a corresponding value in $\Delta N_{\nu}$ given by.
 
\begin{equation}
\biggl(\sum_{i = e, \mu, \tau} \rho_{\nu_i}\biggr) +\rho_{DR} \equiv (3+\Delta N_{\nu})\rho_{\nu_e} \equiv N_{eff} \rho_{\nu_e} ~~,
\label{neffdef}
\end{equation}     
where $\rho_{\nu_i}$ corresponds to the sum over neutrino plus  anti-neutrino energy densities
\begin{equation}
\rho_{\nu_i}  = 2\frac{7}{8}\frac{\pi^{2}}{30}T_{\nu_i}^{4} ~~,
\end{equation}
where $T_{\nu_i}$ is the temperature of each neutrino species.  Note, that since each neutrino species is slightly heated by the $e^+e^-$ annihilation before it decouples at  a different temperature, $\Delta N_\nu = 0.046$ even in the standard big bang.
 
The dark radiation arising from the RS model is different from the other possible "dark" relativistic particles (e.g.  sterile neutrinos). Indeed, during the BBN epoch the dark radiation in the RS model is nearly equivalent to an effective neutrino species. However it acts differently on the CMB.  Whereas light neutrinos or non-interacting particles
can stream and gravitate, a dark radiation term remains uniform everywhere.  Thus, as clarified below, there is a cosmological sensitivity to either relativistic or light  neutrinos at the CMB epoch, particularly given the fact that  their number density is comparable to that of CMB photons.   A dark radiation term of the form of interest here, however, has a different effect on the CMB.

This high density of free streaming particles can inhibit the growth of structure at late times, leading to changes in large scale structure (LSS) that can be constrained by the CMB and matter power spectrum.  In particular, the number of neutrino species primarily affects the CMB by altering the photon diffusion (Silk damping) scale relative to the sound horizon. The sound horizon sets the location of the acoustic peaks while photon diffusion suppresses power at small angular scales. 
This affects both the  ISW effect and the look back time.  Hence, the effect of RS  dark radiation is not equivalent to adding $\Delta N_{\nu }$ neutrino species.  Moreover, if  the added neutrino has a light mass, the CMB constrains that mass through its effect on structure growth in two ways: 1) the early Integrated Sachs Wolfe (ISW) effect, and 2) gravitational lensing of the CMB by LSS.  Indeed, A significant fraction of the power in the CMB on large angular scale is from the early ISW effect, but unaffected by the RS dark radiation of interest here.  Hence, the CMB constraints on the dark radiation discussed here are not equivalent to the constraints on $\Delta N_{\nu }$  deduced in the {\it Planck} cosmological parameters paper \cite{PlanckXIII}.

Although the Planck cosmological parameters paper \cite{PlanckXIII}  mentions �dark radiation�, they use that term in the sense of a non-interacting or massless particle, not as the effect of higher dimensional curvature.   Hence, "dark radiation" was  treated an effective neutrino  species. Similarly, the recent review of Cyburt et al \cite{bbnreview} does not mention dark radiation. Although a number of papers in our scan the literature mention the term �dark radiation�, they invariably refer to an extra relativistic species that can be treated as an effective number of neutrinos, not the effect of curvature in  extra dimension.  Similarly, although there are many references to the extra-dimensional dark radiation term in the literature, there has been no treatment of the combined BBN and CMB constraints since \cite{Ichiki}.  Hence, there is a need to revisit this issue in light of the recent Planck results and revised light element abundance constraints.

\section{BBN constraint}
The light element abundances present at the end of BBN can be used to constrain the physical conditions during the nucleosynthesis epoch. We use a standard BBN code \cite{Smith93} with a number of reaction rates updated \cite{NACRE}. We also replaced the modified Bessel function approximation with a more accurate Fermi-integral solver and checked for any differences in the final abundances. The primordial helium abundance ($Y_p$) is deduced from HII regions in metal poor irregular galaxies extrapolated to zero metallicity \cite{bbnreview}. We adopt  $Y_p = 0.2449 \pm 0.0040 (2\sigma)$ from \cite{bbnreview}.

Deuterium is measured from the spectra of Lyman-alpha absorption systems in the foreground of high redshift QSOs.  A two sigma region of 
\begin{equation}
2.45\times 10^{-5}\leq {\rm D/H} \leq 2.61\times 10^{-5} \hspace{0.5cm}
\end{equation}
is adopted from \cite{bbnreview}. 
 
The inclusion of positive dark radiation implies extra energy density during BBN. It's effects has been studied by many authors \cite{Olive,Yahiro,Orito}. Positive dark radiation increases the cosmic expansion rate  and causes the nuclear reactions to freeze out at a higher temperature. As a result, the neutron to proton ratio increases, since the $n/p$ ratio is related to a simple Boltzmann factor at freezeout ,
\begin{equation}
n/p = \exp{ ({-\Delta m/T})}~~,
\end{equation}
where $\Delta m = 1.293$ MeV is the neutron-proton mass difference, and $T$ is the photon temperature. The increased neutron mass fraction from a positive dark radiation term   increases the D/H and $Y_p$ abundances.   In addition, the faster cosmic expansion results in the freezeout of the deuterium destruction via the reactions $^2$H($d$,$n$)$^3$He and $^2$H($d$,$p$)$^3$H at a higher temperature.  This also leads to a larger deuterium abundance.  The abundances of $^3$H and $^3$He are larger for a positive dark radiation.  This is because these nuclides are mainly produced via the reactions $^2$H($d$,$n$)$^3$He and $^2$H($d$,$p$)$^3$H, respectively, and the deuterium abundance is higher.  When the dark radiation is negative the opposite effect occurs.

The primordial $^7$Li abundance is deduced from the observed abundances in low mass metal poor stars. 
For $^7$Li we adopt the $2 \sigma$ constraint of \cite{bbnreview}
\begin{equation}
1.00\times 10^{-10}\leq {\rm ^7Li/H} \leq 2.20\times 10^{-10} ~~.
\end{equation}
\par
There is a well known lithium problem \cite{bbnreview} whereby the predicted primordial lithium abundance exceeds the observed primordial lithium abundance by about a factor of 3 for $\eta = (6.10 \pm 0.04) \times 10^{-10}$ deduced from the {\it Planck} analysis \cite{PlanckXIII} of the CMB primordial power spectrum.

\begin{figure}[t]
 \includegraphics[scale=0.40]{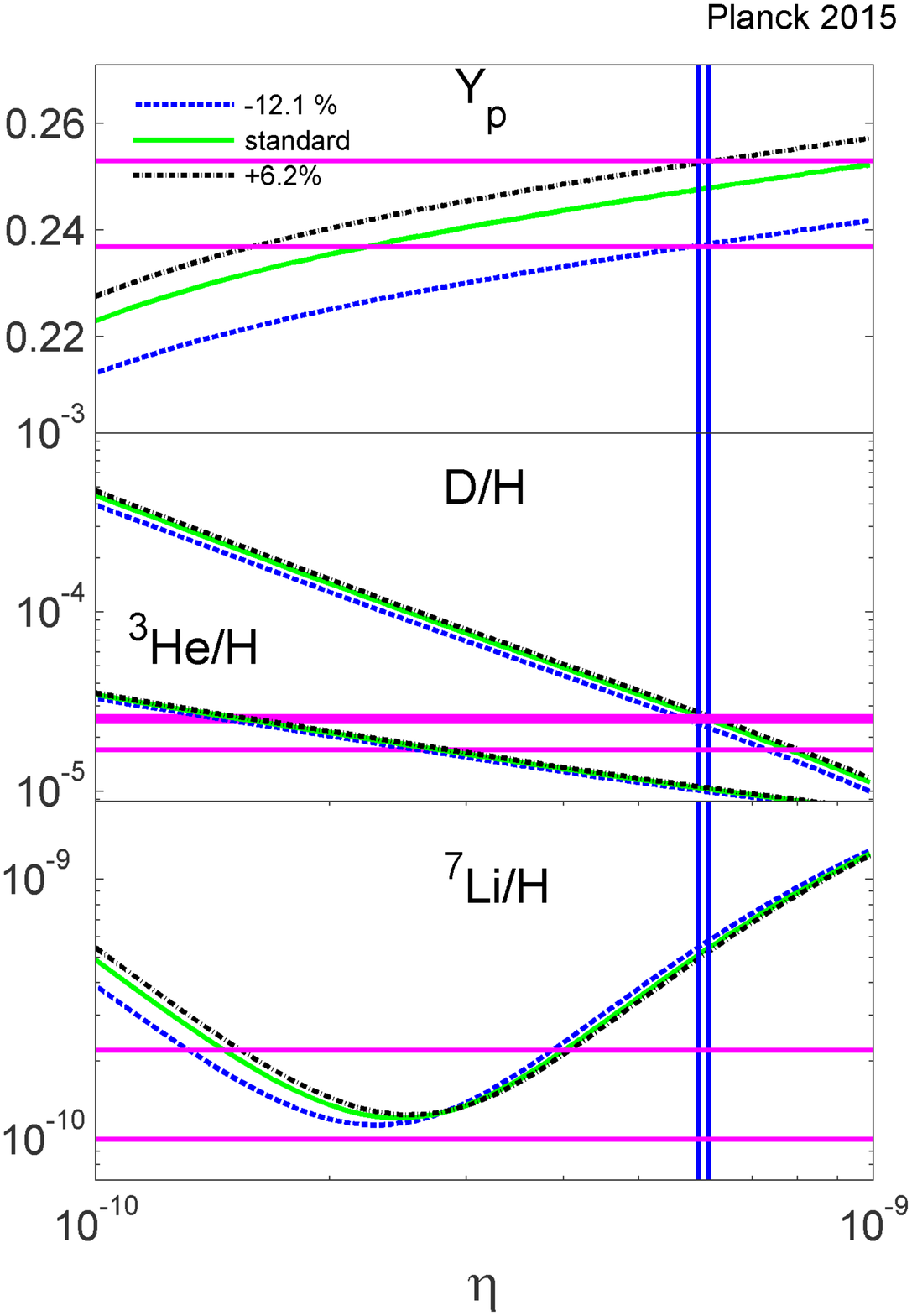}
 \caption{(Color online) Light element abundances, $Y_p$, D/H, $^3$He/H, and Li/H as a function of baryon to photon ratio $\eta$. The red horizontal lines correspond to the adopted observational upper and lower limits on primordial abundances. The solid green line is the result for the standard BBN with no dark radiation. The dot dashed black line  and the dashed blue line show the results of BBN in which the energy densities in dark radiation are $+6.2\%$ and $-12.1\%$, respectively, of the total relativistic particle energy density (at 10 MeV). The two lines correspond to  upper and lower limits on $\rho_{DR}$ derived from the light element abundances.  The vertical solid blue lines show the CMB constraint on $\eta$ from {\it Planck} \cite{PlanckXIII}.  
}
  \label{fig:1}
       \end{figure}

Figure \ref{fig:1} shows the calculated light element abundances, $Y_p$, D/H, $^3$He/H, and Li/H as a function of $\eta$. The solid green line is the result for the standard BBN with no dark radiation.  The dot dashed black line and the dashed blue line show the results of BBN in which the energy densities of the dark radiation are $+6.2\%$ and $-12.1\%$, respectively, of the total particle energy density. The two lines correspond to the cases of the upper and lower limits on $\rho_{DR}$ derived from the constraints on light element abundances.  The vertical solid blue lines enclose the $\pm 1\sigma$ constraint on $\eta$ \cite{PlanckXIII}.  The horizontal lines correspond to the observational upper and lower limits on primordial abundances.

For the case of  positive dark radiation, we find a decrease of the $^7$Li abundance for $\eta \lesssim 3\times 10^{-10}$ and an increase for $\eta \gtrsim 3\times 10^{-10}$.  We note that the primordial $^7$Li nuclei are produced as $^7$Li in the low $\eta$ region and $^7$Be in the high $\eta$ region during BBN. A positive dark radiation term leads to a slight excess of the $^7$Li abundance because $^7$Li is produced via the $^4$He($t$,$\gamma$)$^7$Li reaction and the abundance of $^3$H is higher. There is also less time for the lithium destruction reaction $^7$Li$(p,\alpha)^4$He. On the other hand,  a positive dark radiation decreases the $^7$Be abundance. The slight increase in the $^3$He abundance results in a somewhat increased production rate of $^7$Be via the reaction $\he4 (\he3,\gamma)\be7$. However, the significant increase of the neutron abundance leads to an enhanced destruction rate of $^7$Be via the reaction $^7$Be($n$,$p$)$^7$Li. As a result, the final $^7$Be abundance decreases. In either case, dark radiation does not affect the primordial lithium abundance sufficiently to solve the lithium problem without violating the $^4$He and deuterium constraints.  Hence, we presume that the lithium problem is solved by another means and do not utilize the $^7$Li abundance as a constraint on dark radiation.

We estimate the likelihood for $\rho_{DR}/\rho$ assuming a gaussian prior on the observational limits of D/H and $Y_{p}$. We define the marginalized likelihood function by
 
\begin{equation}
L(\rho_{DR/}/\rho)= \int_{\eta}L_{D/H}L_{Y_p} d\eta
\end{equation}
where
\begin{equation}
  L_{i}=\frac{1}{\sqrt{2\pi}\sigma_{i}}\exp
    \left\{-\frac{\left[Y_{i,{\rm BBN}}(\rho_{DR}/\rho, \eta)-Y_{i,{\rm obs}} \right]^2}
    {2\sigma_{i}^2}\right\}
\end{equation}

\begin{figure}[H]    
    \includegraphics[width=3.5 in]{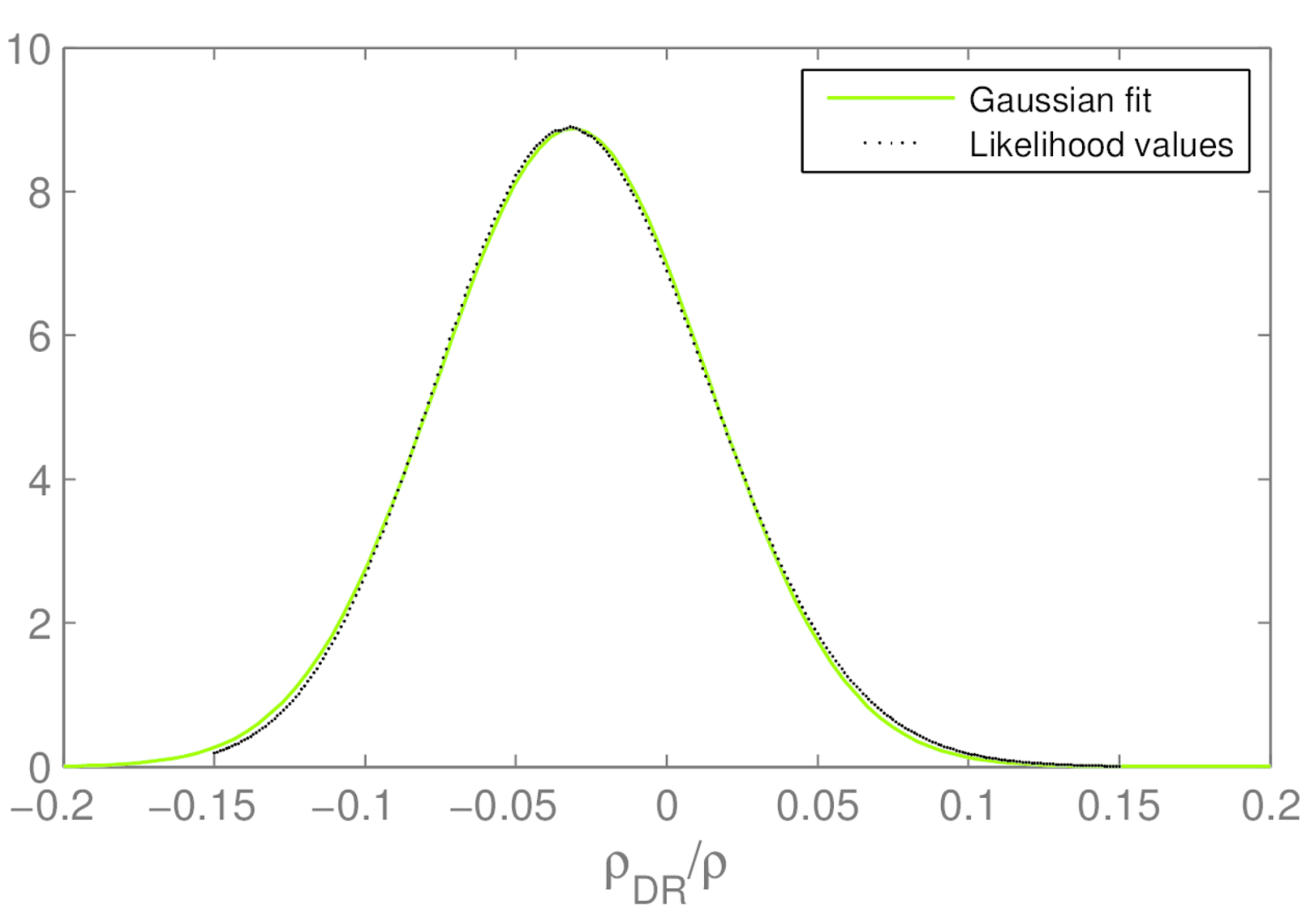}
    \caption{Likelihood function for $\rho_{DR}/\rho$(black line), with a Gaussian distribution (green line) }
    \label{fig:2}
\end{figure}

The $L_{D}$ and $L_{Y_p}$ are the likelihood values we obtain from the deuterium and helium abundances. $Y_{i,BBN}$ is the yield calculated by the BBN code and $Y_{i,obs}$ is the observational abundance. $\sigma_{i}$ is the uncertainty of the observational abundances. Here $i$ can be either ${\rm D/H}$ or $Y_{p}$. A plot of the Gaussian fit to the likelihood values for $\rho_{DR}/\rho$ is given in Figure \ref{fig:2}.

We obtain 1-$\sigma$ bound on $\rho_{DR}/\rho$ of (-3.10 $\pm$ 4.49 \%). This corresponds to  $N_{eff}$ in the range of 2.81 $\pm$ 0.28, or in terms of an equivalent number of neutrino species, via Eq.~(\ref{neffdef}) we have $\Delta N_\nu = -0.19 \pm 0.28$. This is comparable to the BBN+$Y_p$+D value of $N_{eff} = 2.85 \pm 0.28$ deduced in Ref.~\cite{bbnreview}.

\section{CMB constraints}
Although the epoch of photon last scattering is in the matter dominated epoch, there remains an effect on the CMB power spectrum due to the still significant contribution from relativistic mass energy and the effect of the uniform dark radiation term on the expansion rate and acoustic oscillations of the cosmic fluid. On the other hand, the CMB power spectrum is very sensitive to a number of other  parameters that have little or  no effect on BBN. Thus, to obtain a total constraint on the dark radiation contribution to energy density, we have performed a simultaneous fit to the {\it TT} power spectrum of temperature fluctuations in the CMB. To achieve this, we have fixed most of the cosmological parameters to their optimum values \cite{PlanckXIII} and only varied the  dark radiation content and $\eta$ in the fit. 
Fits were made to the  {\it Planck} data \cite{PlanckXIII} using the CAMB code \cite{Camb}.  In the limit of no dark radiation we recover the {\it Planck} value of  $\eta =  (6.10\pm 0.04) \times 10^{-10}~(1 \sigma)$ \cite{PlanckXIII}.  For $\eta$ fixed by the {\it Planck} analysis, the $2 \sigma$ constraint from the CMB alone would imply $-9.0$ \% $< \rho_{\rm DR} /\rho < 13.5$ \%.

We note, however,  that the deduced dark radiation content is sensitive to the adopted value of $H_0$.  In the present work we utilize  $H_0 = 66.93 \pm 0.62$ km s$^{-1}$ Mpc$^{-1}$ (Planck+BAO+SN) from the {\it Planck} analysis \cite{PlanckXIII}.   However, a larger value is preferred \cite{Riess16} from local measurements of $H_0$, and a larger value  of $H_0 = 73.24 \pm 1.74$ km s$^{-1}$ Mpc$^{-1}$ was  obtained \cite{PlanckXIII}  when adding a prior on $H_0$.   Adopting this larger value would shift the inferred dark radiation constraint toward larger positive values.  We prefer the lower value of $H_0$ deduced by {\it Planck} because this discrepancy between the local value and the CMB value would in fact be explained \cite{Riess16} by  the presence of dark radiation at the CMB epoch.

\begin{figure}[h]
\includegraphics[width=3.5 in]{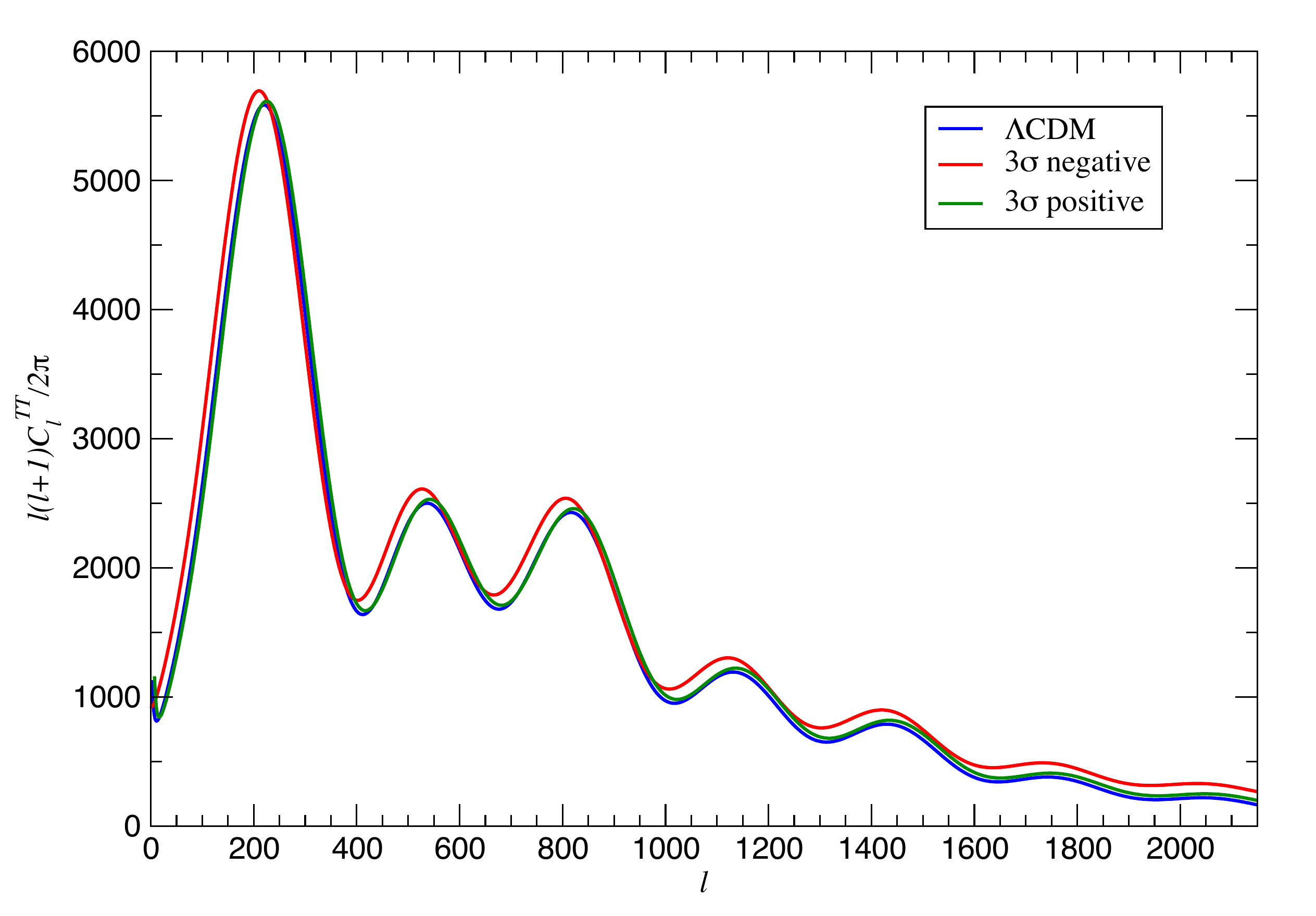} 
\includegraphics[width=3.5 in]{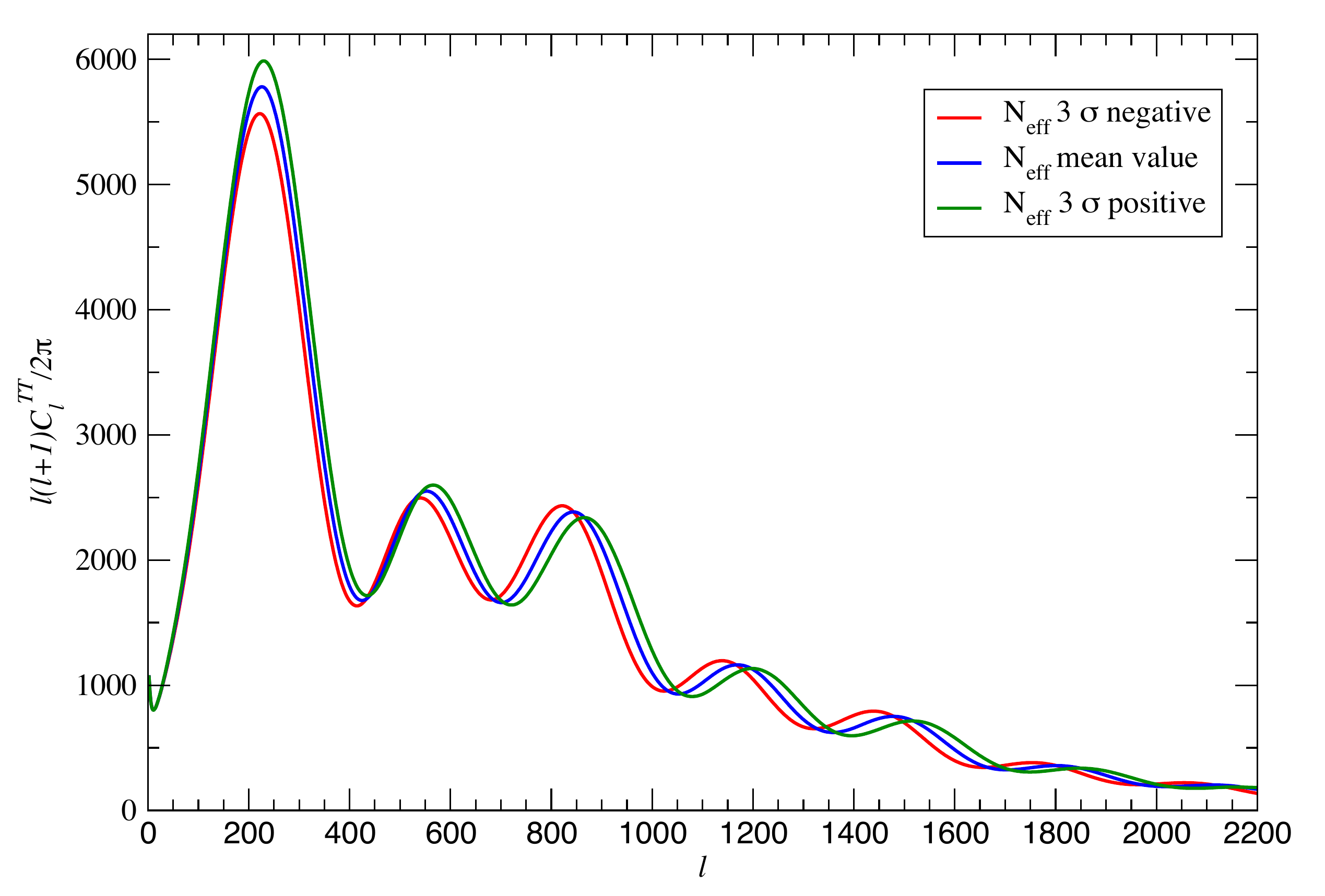} 
\caption{(Color online) Upper panel illustrates the effect of Dark Radiation on fits to the {\it TT} CMB power spectrum.  Lines drawn indicate the $\pm 3\sigma$ deviations in the BBN constraint as labeled. Lower panel shows the equivalent BBN $\pm 3\sigma$ constraints, but in this case it is treated as an effective number of neutrino species.  Clearly the effect of dark radiation on the CMB is different than an equivalent number of neutrino species.}
\label{fig:3}
\end{figure}

 It is important to appreciate that adding a dark radiation term is not equivalent to adding an effective number of neutrino species to the CMB analysis.  This is illustrated in Figure \ref{fig:3}.  The upper and lower panels of Figure \ref{fig:3} show the effects on the CMB {\it TT}  power spectrum of adding dark radiation vs. an effective number of neutrino species, respectively.  This figure plots the usual normalized amplitude $C_l$ of the multipole expansion for the TT power spectrum as a function of the moment $l$.  As can be seen on the upper  figure, a positive dark radiation has only a slight  effect, while a negative  dark radiation term (red line) slightly increases the amplitude of the acoustic peaks due to the diminished expansion rate.  However, a relativistic neutrino-like species can stream and gravitate.  Therefore, it has the opposite effect of increasing the  amplitude of the first acoustic peak for a positive contribution while decreasing the amplitude for a negative contribution.  In addition, a relativistic species  also shifts  the  location of the higher harmonics.  Thus, it is important to re-examine the CMB constraints on the  brane-world dark radiation term independently of any previously derived  constraints on the effective number of neutrino species.

\begin{figure}[htb]
\includegraphics[width=3.5 in]{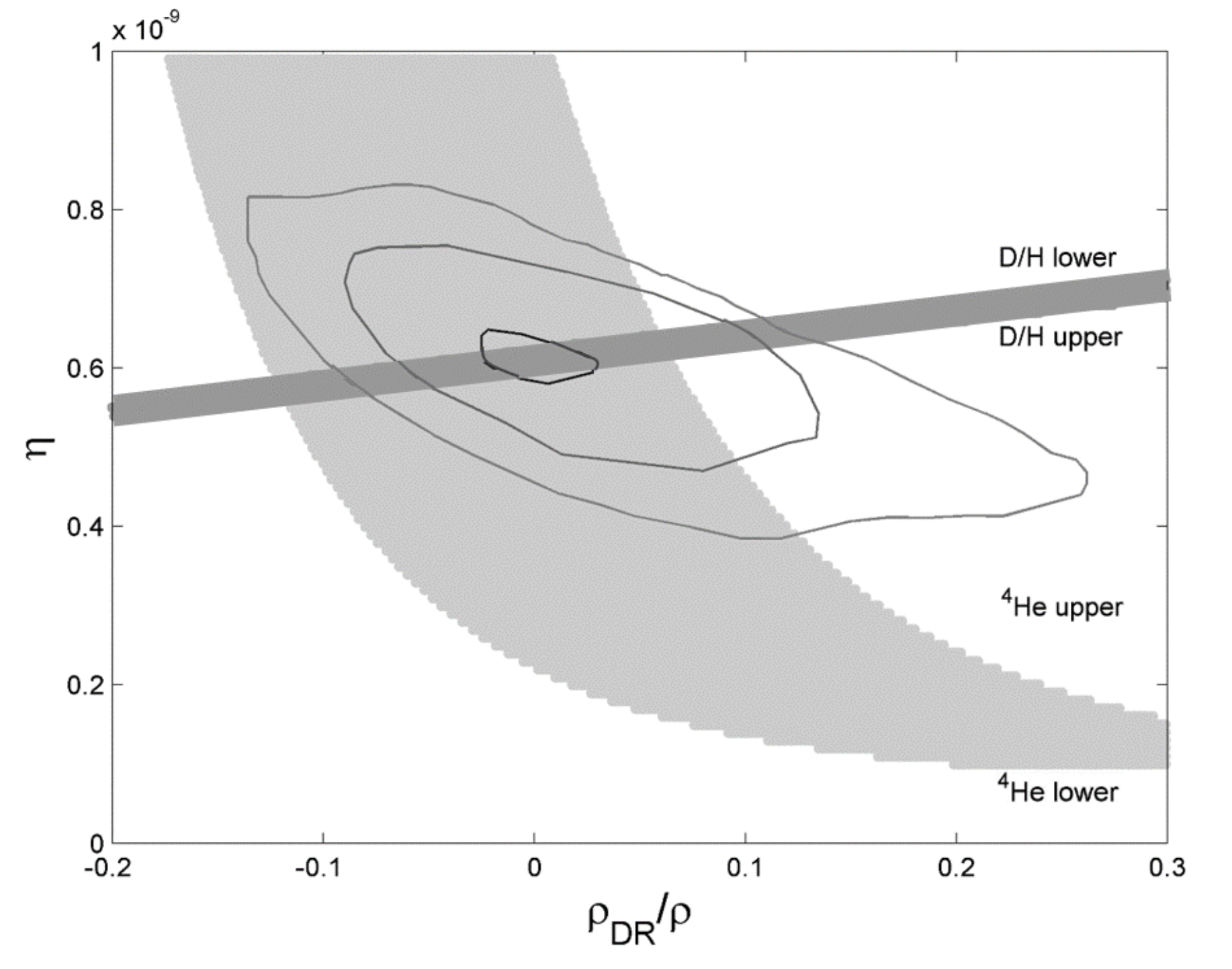} 
\caption{(Color online) Constraints on  dark radiation in the $\rho_{\rm DR} /\rho$ vs.~$\eta$  plane.  Contour lines  show the $1,~ 2, $  and $3 \sigma$ confidence limits  based upon our fits to the CMB power spectrum.    Dark shaded lines show the constraints from the primordial deuterium abundance as described in the text.  The light shaded region shows the $Y_p$ constraint.}
\label{fig:4}
\end{figure}

 Figure \ref{fig:4} shows the combined constraints on $\eta$ vs.~dark radiation based upon our fits to both BBN and the CMB power spectrum.  The  contour lines on Figure \ref{fig:4} show the CMB $1,~ 2,$  and $3 \sigma$ confidence limits in the $\eta$ vs.~dark radiation plane. The shaded regions show the BBN $Y_p$ and D/H constraints as labeled. 

The best fit concordance shown in Figure \ref{fig:4} is consistent with no dark radiation although in the BBN analysis there is  a slight  preference for  negative dark radiation.  A similar result was found in  the previous analysis of Ref.~\cite{Ichiki}.  However, the magnitude of any dark radiation is much more constrained in the present analysis.  This can be traced to both the CMB and new light-element abundances.
Also, it is worth mentioning that the value of $\eta$ deduced from the WMAP data is $(6.19\pm 0.14)\times 10^{-10}$, while the (WMAP+ BAO + H$_0$) data is $(6.079\pm 0.09)\times 10^{-10}$  \cite{WMAP9}.  Hence, the BBN dark radiation constraint based upon   the WMAP 
results  would be nearly identical and would also have a slight  preference for a negative dark radiation.

We note, however, that the results in Figure \ref{fig:4} are sensitive to the  the value of $Y_P = 0.2449 \pm 0.0040$ adopted from Ref.~\cite{bbnreview} based upon  the recent primordial helium abundance determinations from Aver, Olive, and Skillman \cite{Aver15}.   This result is based on data from Izotov and collaborators \cite{Izotov14}.   However, using data that overlaps strongly with  that of Ref.~\cite{Aver15}, a higher helium abundance of $Y_P = 0.2551 \pm 0.0022$ was deduced in \cite{Izotov14}. If we adopt the larger value for the helium abundance then the $2\sigma$  
BBN constraint on the dark radiation content increases from to positive values of $+3.2$ \% $< \rho_{\rm DR} /\rho < 14.4$ \%.  Nevertheless we prefer the Bayesian analysis of \cite{Aver15} as it arguably incorporates  a better treatment of correlated errors.   It is also more consistent with the CMB constraint.

\section{Conclusion}
 In conclusion we deduce  that based upon our adopted  $2 \sigma$ (95\% C. L.) BBN constraints, brane-world dark radiation is allowed  in the range of $-12.1\%$  to  $+6.2\%$ ($\Delta N_\nu = -0.19 \pm 0.56$) compared to the range deduced in Ref.~\cite{Ichiki} of  $-123\%$ to $10.5\%$ based upon constraints available at the time of that paper. 
After taking into account the $2\sigma$ limits on  the dark radiation from the fit to the CMB power spectrum, this region shrinks to a range of $-6.0\%$  to $+6.2\%$ ($\Delta N_\nu = -0.19 _{-0.18}^{+0.56}$).  However, if the higher helium abundance of \cite{Izotov14} were adopted, the allowed range increases to 
The $1 \sigma$ BBN constraint ($2.63 \leq N_{\nu} \leq 3.38$, $\Delta N_\nu = -0.19 \pm 0.28$) is comparable to the values deduced by \cite{bbnreview,PlanckXIII}. 
For $\eta$ fixed by the {\it Planck} analysis, the constraint on positive dark radiation comes from the upper bound on the $^4$He mass fraction and the upper bounds on the D/H. The limit on negative dark radiation arises from the constraint on cosmic expansion rate at the epoch of last scattering (the CMB) and the lower bound of D/H.  

We caution, however, that either a larger value for the Hubble parameter \cite{Riess16} could shift the allowed CMB range (or a larger primordial helium abundance \cite{Izotov14} could shift the  BBN range) to a higher positive contribution of dark radiation.
For example,  if the higher helium abundance of \cite{Izotov14} were adopted, the allowed range increases to  $+3.2$ \% $< \rho_{\rm DR} /\rho < 13.5$ \%.  In this case the lower bound is from BBN and the upper bound is from the CMB.

We also checked the corresponding $^{7}$Li/H abundance for the allowed ranges of dark radiation. For dark radiation of $+6.2\%$ and $\eta$ of $6.1 \times 10^{-10}$  the lithium abundance is $5.19 \times 10^{-10}$ slightly alleviating the lithium problem. The $3\sigma$ CMB contour combined with D/H corresponds to a lower limit of dark radiation of $-9.1\%$. In this case the corresponding lithium abundance is increased to $5.64 \times 10^{-10}$ exacerbating the lithium problem. Hence, although  a positive dark radiation slightly reduces the lithium abundance, it is not sufficient to solve the lithium problem.

\begin{acknowledgements}
Work at the University of Notre Dame is supported by the U.S. Department of Energy under Nuclear Theory Grant DE-FG02-95-ER40934. One of the authors (MK) acknowledges support from the
JSPS.

\end{acknowledgements}

\end{document}